# Quantum behavior of graphene transistors near the scaling limit

Yanqing Wu[†*], Vasili Perebeinos[†], Yu-ming Lin, Tony Low, Fengnian Xia and Phaedon Avouris[*]

*IBM Thomas J. Watson Research Center, Yorktown Heights, NY 10598*

**Abstract**. **The superior intrinsic properties of graphene have been a key research focus for the past few years. However, external components, such as metallic contacts, serve not only as essential probing elements, but also give rise to an effective electron cavity, which can form the basis for new quantum devices. In previous studies, quantum interference effects were demonstrated in graphene heterojunctions formed by a top gate. Here phase coherent transport behavior is demonstrated in a simple two terminal graphene structure with clearly-resolved Fabry-Perot oscillations in sub-100 nm devices. By aggressively scaling the channel length down to 50 nm, we study the evolution of the graphene transistor from the channel-dominated diffusive regime to the contact-dominated ballistic regime. Key issues such as the current asymmetry, the question of Fermi level pinning by the contacts, the graphene screening determining the heterojunction barrier width, the scaling of minimum conductivity and of the on/off current ratio, are investigated.**

[†]These authors contributed equally to this work.

[*]Email: ywu@us.ibm.com (Y.W.) and avouris@us.ibm.com (P.A.)

Graphene is a zero-bandgap, two dimensional material with ultra-high carrier mobility and velocity[1-2], properties which are highly desirable in high-speed and radio-frequency (RF) applications[3-5]. Electrons in graphene behave as Dirac fermions, exhibiting many unique transport phenomena such as Klein tunneling and suppression of carrier backscattering[6-7]. In particular, graphene heterojunctions are being proposed for novel applications involving electron manipulation in two-dimensions, such as electron focusing, collimation and guiding.[8-11]. Previously, the best available graphene heterojunction structures were demonstrated using ultra-short top gate structures[12-15], whose fabrication usually involves rather complex and non-scalable fabrication processes and whose properties are sensitive to the top-gate oxide quality and thickness. Here we demonstrate that a simple, two-terminal device formed by standard lithography, where the heterojunction is formed by metal contact-induced doping instead of a top-gate, can exhibit quantum interference oscillations.

In graphene research, most transport studies have been carried out using Hall-bar structures to exclude metallic contact-induced effects and to probe the intrinsic electronic properties. Indeed, very high mobilities have been attained by reducing external perturbations[16-17]. However, the source and drain metallic contacts alter the intrinsic graphene properties and produce an effective electron resonance cavity which allows us to study quantum effects that are unique to graphene. The full understanding of the entire graphene device including the source and drain electrodes is of vital importance in technology, especially when the contacts dominate electrical transport. In this work, by aggressively scaling the channel length to the point where the transport is ballistic and phase coherent, we are able to probe the quantum transport behavior in graphene transistors. This allows us to address various key issues such as Fermi-level pinning of graphene underneath the metal, metal induced heterojunction barriers, non-linear screening, and the origin of minimum conductivity. Unlike other conventional semiconductors, all of these properties are uniquely determined by the graphene electronic structure. Thus, our study allows us to obtain essential information needed to explore quantum transport in ultra-small graphene devices, as required in the development of practical graphene transistors.

**Transport mechanism of a back-gated graphene transistor**

Arrays of transistors on mechanically exfoliated graphene were fabricated on a 90 nm silicon dioxide ($SiO_2$) film on top of a heavily p-doped silicon substrate that serves as a bottom gate. The metal contacts for source and drain consist of a metal stack of 20 nm Pd and 30 nm Au with the Pd layer in direct contact with graphene. The transistors have the same channel width $W = 1\mu m$ and varying source-drain spacing, which defines the channel length, varying from 0.5 $\mu$m down to 50 nm, as shown in the scanning electron microscopy (SEM) image of Fig. 1a. The smallest graphene transistor with channel length of 50 nm is shown in the inset of Fig. 1a. For electrons injected from the source region, their transport across the device can be analyzed as involving a series of five distinct regions, as shown schematically in Fig. 1b. Carriers first need to pass through a junction formed at the interface of the metal and the graphene underneath it (region I), with a certain transmission efficiency denoted as $T_{MG}$. Since the graphene underneath the metal is doped by charge transfer (p-type in this case), a p-n or p-p' junction forms between the graphene underneath the metal and the graphene in the channel region in the vicinity of the metal edges (regions II and IV). The contact junction width is not determined in this case by the oxide thickness alone, but rather by electrostatic non-linear screening[18-19]. In conventional transistors under sub-threshold conditions with negligible carrier density, the structure of the devices, such as gate oxide and the channel body thickness, determine the screening length. In the case of graphene, the carrier density always depends non-linearly on the electrostatic potential, so that a new carrier density dependent lenghscale[18-19] enters into the Poisson problem, namely $l_M = 4\kappa\varepsilon_0\hbar v_F / \left(\sqrt{\pi n_M}\, e^2\right)$

(see methods). In graphene transistors, for typical carrier densities $n_M$ and dielectric environment $\kappa$, $l_M$ should not exceed a few nm. If this is true, the junction width is expected to be smaller than the mean free path $\lambda_{\text{mfp}}$ even in devices with a large gate dielectric thickness of 90 nm as in our case. Region III involves transport in the main graphene channel with length $L_{\text{ch}}$, where the carriers can undergo scattering and may lose their ballisticity depending on the ratio of $\lambda_{\text{mfp}}$ and $L_{\text{ch}}$. The graphene junction at the drain side has the opposite polarity of that at the source side and the overall structure can thus be viewed as two back-to-back p-n diodes. If the channel length is short enough so that transport across the channel remains ballistic, the back-to-back junctions form a ballistic p-n-p or p-p'-p structure depending on the back gate voltage. In region V, where carriers pass from graphene underneath the metal to the drain electrode, the process is identical to that in region I. Typically, symmetric electric currents for electron and hole branches are observed in measurements of Hall-bar graphene structures, as is expected from the comparable electron and hole mobilities in graphene. In a two-terminal configuration, however, as will be discussed further in this work, asymmetric electric currents for electron and hole branches are usually observed and attributed to contact-induced doping[20-25]. In a long channel device where symmetric diffusive transport in the channel region dominates, the effect of contact doping can be studied by measuring the odd part of the total device resistance, $R_{\text{odd}} = (R_{\text{n}} - R_{\text{p}})/R_{\text{p}})$ [26-27]. It is noted here that the odd resistance, which averages the contributions from carriers injected through different angles, is sometimes used as indirect evidence of Klein tunneling through the doped junctions[14-15]. Fig. 1c and Fig. 1d show the characteristic ambipolar gate-dependent transport of graphene transistors with three representative channel lengths at room temperature and 4.3 K, respectively. It is immediately evident that as channel length decreases, the asymmetric contribution to resistance due to contact-induced junction increases relative to the total device resistance. Specifically, the p-side resistance decreases faster than the n-side resistance. This behavior and other transport phenomena due to channel length scaling are systematically analyzed below.

**Quantum interference in a two-terminal graphene device**

The quasi-ballistic transport of graphene devices as a function of channel length is described using Landauer's approach. The conductance is given by the number of the participating conduction modes $M$ and an averaged transmission coefficient $T_K$ as: $G = (4e^2/h)MT_K$ [28]. The number of modes in the absence of charge puddles and at zero temperature is given by $M = (\Delta E_F/\pi\hbar v_F)W$, where $\Delta E_F$ is the Fermi level shift relative to the Dirac point and $v_F$ is the Fermi velocity ($\sim 1\times10^8$ cm/s). On the other hand, diffusive transport is described by the Boltzmann-Drude conductivity $\sigma = en\mu$. The two approaches give the same result for a particular choice of the mean free path[29]. The 2D $\lambda_{mfp} = \frac{\pi}{2}v_F\tau$ is determined by the scattering time $\tau$. From the mobility $\mu = 4e\lambda_{mfp}/h\sqrt{\pi n}$ measured at carrier density $n = 6 \times 10^{12}$ cm$^{-2}$, the mean free path in our two-terminal device is found to be about 80 nm at low temperature and 60 nm at room temperature. Therefore, in devices with channel length around 50 nm, transport is expected to be quasi-ballistic. Since the elastic mean free path is typically shorter than the phase coherence length in graphene at low temperatures, coherent transport across the small channel devices is expected. The transport processes involve carrier injection from a metallic contact and traveling through the two potential barriers (regions II and IV). Although most of the channel region is controlled solely by the back gate voltage, the two junction regions (region II and IV) in the vicinity of the metal contacts are already doped by the metal and thus the carriers there may have a different polarity than that of the carriers in the channel. The resulting heterojunction can be modeled as two bipolar junctions, which form a resonant electron cavity that can lead to quantum interference. Indeed, in the device with channel length of 50 nm, where transport is quasi-ballistic and phase-coherent, clearly-resolved current oscillations are observed at the n-branch as shown by the blue curve in Fig. 2a. No such oscillations are observed in the p-branch. The distinctly different behavior of the electron and hole branches confirms that the graphene was p-doped underneath the metal contacts by charge transfer from the metal. In the p-branch, the interference of holes is much weaker due to the nearly transparent heterojunctions formed in a p'-p-p' structure[7,30-31].

**Modeling of an ideal graphene heterojunction**

To analyze the oscillations seen in the experiment in Fig. 2a, we first consider a simple model (1) for the device resistance in the ballistic limit with transparent contacts, i.e. with $T_{MG}=1$ and a pinned Fermi level in graphene under the metal. For the electrostatic potential, which controls the transmission $T_K$ through the heterojunction and, hence, the amplitude of the oscillations, we use the model described in the Methods section. The electrostatic potential profile of the graphene heterojunction induced by the contacts was predicted to decay fast, within a range of a few nanometers[18-19]. The width of this barrier, defined here as the position at half maximum, is mainly determined by the effective dielectric constant of graphene on the substrate, which in this case is $SiO_2$ (see the Supplementary information). The resistance of the 50 nm device, shown as the red dashed curve in Fig. 2a, is computed using the Transfer Matrix Method (TMM)[32] with essentially one varying parameter in the model, the Fermi level of graphene under the metal, which is first assumed to be pinned, $V_0 = 100$ meV. This choice of $V_0$ is consistent with theoretical predictions[21] and the measured gate dependence of the resistance on the p-branch, which can be reproduced by a more sophisticated simulation using the same value of $V_0$ (see below). As shown in Fig. 2a, the period of the oscillations is determined by the cavity length in the simulation and it matches well the period observed in the experiment, confirming their nature as electron Fabry-Perot oscillations. The origin of the oscillations can be understood as a coherent tunneling and reflection between two barriers with the transmission coefficient $T\left(p_y\right)$ : $G = (4e^2/h)W\int_{-\infty}^{\infty}T\left(p_y\right)dp_y/2\pi$ , with $T\left(p_y\right) = \dfrac{T^2}{\left|1 - R\,\mathrm{e}^{i\phi}\right|^2}$ , where $p_y$ is the transverse electron kinetic momentum with the phase shift gained in a roundtrip $\phi = 2k_x L_{ch} + \chi$ determining the maximum resistance with $\chi$ the phase shift from the tunneling of the two barriers[7,30-31], each with transmission $T$ ($R=1$-$T$). Both experimental resistance peak positions and those from the TMM simulations agree well with the simple interferometer expression $E_F \propto \dfrac{\hbar v_F}{L_{cav}}\pi N$ as shown in Fig. 2b using as a single fitting parameter the cavity length $L_{cav}$, The experimental Fermi levels were found from the gate voltage $V_{BG}$ and back gate capacitance $C_{BG}$ using the relation $E_F = \hbar v_F\sqrt{\pi C_{BG}V_{BG}}$ . Agreement between the experimental results and the interferometer model is obtained when $L_{cav} \approx 50$ nm, i.e. when the cavity length is very close to the measured channel length.

The small barrier width deduced in this way verifies the theoretical predictions[18-19,24] that non-linear screening effects play a very important role in graphene. However, while the oscillation patterns in the ballistic simulation coincide with the experimental ones, the overall gate dependence of the simulated resistance deviates significantly from the experimentally observed behavior. In particular, the resistance asymmetry predicted by our simulations is only about 30%, while the measured data suggest a factor of three stronger asymmetry.

Although, by assuming a larger barrier width one could obtain a stronger asymmetry, the larger width would also lead to a period larger than that experimentally observed in Fig. 2a. (see Supplementary information Fig. S3). Therefore, a more realistic contact model is required in order to simultaneously reconcile both of these experimental features, as discussed next.

**Modeling of a realistic graphene heterojunction**

To describe the overall shape of the experimental resistance quantitatively, we extended the model to allow for (a) the modulation of the graphene Fermi level under the metal by the back gate voltage, i.e. not assuming Fermi level pinning, (b) the broadening of the electronic states under the metal, and (c) the presence of electron-hole puddles in the channel. The effect of the contacts is captured by rescaling the ideal heterojunction resistance $R_K$ according to $R_{tot} = R_K \left( 2/T_{MG} - 1 \right)$. We chose realistic values for the parameters in model (2), specifically, the metal-induced doping $V_0$, the effective graphene-metal electrostatic distance $d_1$, which determines the capacitive coupling to the metal[32-33] (see Supplementary information), the graphene–metal contact transparency $T_{MG}$, the electron-hole puddle density in the channel $n_{pd}$, and the broadening of the graphene DOS underneath the metal $V_{pdM}$. (See Supplementary information ) The result of the simulation is plotted in Fig. 2a as the solid red line. The simulation now agrees remarkably well with the experimental data (blue line), not only for the oscillatory part with well fitted peaks and valleys, but also with respect to the value of the resistance across the entire gate voltage range. The previously underestimated asymmetry between the electron and hole branches is now accounted for by finding in the simulation that a broadened second resistance peak develops, which is associated with the graphene Fermi level crossing the Dirac point underneath the metal at a back-gate voltage around $V_{GD2}$=15-20V. This shows that the partial pinning of the Fermi-level of the graphene underneath the metal can have a strong impact on the transport asymmetry which is typically observed in graphene devices such as those in Figs. 1c and 1d. Using the same model and the same parameters, the ballistic prediction for the total resistance of the same 50 nm graphene device at room temperature also fits the experimental data extremely well, as shown in Fig. 2c, corroborating the validity of our model.

To further explore the chiral behavior of the carriers in a ballistic graphene heterojunction, we applied a magnetic field ($B$), which bends the carrier trajectories and changes the incident angles at the p-n interfaces. The $y$-component of electron kinetic momentum changes to $\tilde{p}_y = p_y - eBx$. Since the non-normal incidence carriers contribute to the oscillatory resistance, the applied magnetic field would result in a phase shift of the quantum oscillations referred to as "half-period" shift at a certain critical $B$-field that depends on the back gate bias[14-15,31]. We investigated the $B$ field dependence of the oscillatory part of the conductance of a longer channel device with the length of 70 nm, which is shown as solid lines in Fig. 2d. The corresponding theoretical predictions shown in Fig. 2d as dashed lines are in good agreement using a fitting parameter of $L_{ch}$ = 70 nm. Both curves display the near π-phase shift as indicated by the gray bars in Fig. 2d when the B-field is changed from 0 T to 2 T (critical field ~ 0.24 T [7]), which again serves as strong evidence of the chiral nature of electrons in graphene and the formation of Fabry-Perot quantum oscillations in this simple two-terminal system. The fact that the experimental resistance peak positions in both the 50 nm and 70 nm devices agree well with the expectations of the interferometer model using cavity lengths close to the geometric length of transistor channels sets the upper-limit on the width of the p-n junction barriers to only a few nanometers, verifying the theoretical predictions[18-19]. Compared to previously used, ultra-short, top-gated resonance cavities, the simple device configuration and fabrication processes employed here present an advantage in potential quantum interference applications.

**Channel length scaling and Fermi level pinning**

The evolution of the ballistic device resistance features as a function of channel length, such as electron-hole asymmetry, minimum conductivity, and the on/off ratio were systematically studied here and the results are summarized in Fig. 3. The electron-hole asymmetry in two-terminal graphene devices, expressed as $(R_n-R_p)/R_p$, is plotted against the $L_{ch}$ for a number of devices both at room temperature and at 4.3 K in Fig. 3a. In this case, $R_n$ and $R_p$ are defined as the resistance at $V_{BG}$-$V_{CNP}$ = 25 V and -25 V, respectively. Here, the charge neutrality point ($V_{CNP}$) is the gate voltage at the maximum resistance. The effective total, incoherent transmission coefficient through the graphene channel, two graphene-metal and two graphene under the metal - graphene in the channel junctions in the Landauer formulation is given by

[28]: $\dfrac{1-T}{T} = 2\dfrac{1-T_{MG}}{T_{MG}} + 2\dfrac{1-T_K}{T_K} + \dfrac{L_{ch}}{\lambda_{mfp}}$, where $T_{MG}$ is a gate independent transmission

coefficient and $T_K$ is a gate dependent transmission coefficient of the p-n junction through Klein tunneling. We note here that $T_K$ is an averaged coefficient for carriers injecting from all angles. In the channel-limited transport regime, the carrier density in the channel determines the device resistance. On the other hand, in the contact-limited transport regime, the number of conduction modes is determined by the carrier density in graphene under the metal with its initial value determined by the metal-graphene charge transfer and further modified by the back gate voltage. The evolution of the asymmetry is clearly channel length dependent as transport evolves from diffusive to ballistic in Fig. 3a. In the case of fully ballistic transport with $\lambda_{mfp}$ much larger than $L_{ch}$, the asymmetry will only be determined by the difference of chiral tunneling efficiency, provided the number of conduction modes is kept the same. At negative gate biases, the channel region is p-type, therefore, the graphene junction is of the p'-p type where $T_K \approx 1$. However, when the back gate is biased positive, the channel region is n-type doped and the graphene junction is of the p'-n type, where $T_K < 1$ because of the scattering of non-normally incident carriers. However, the difference of transmission coefficient of Klein tunneling alone cannot yield the amount of the asymmetry in the channel resistance, as suggested by the fitting in Figs. 2a and 2c, so the changing number of conducting modes of the graphene underneath the metal contributes to the asymmetry significantly. When the back gate switches from negative to positive bias, it not only increases the energy of the Fermi-level in the graphene channel, but also raises the Fermi-level in the graphene underneath the metal so it becomes more weakly p-type, having a decreased number of conduction modes. The strong asymmetry observed experimentally, as a result of these two effects, supports a variation of the Fermi-level of graphene underneath the metal as a function of gate bias. The larger asymmetry observed at 4.3 K is due to the thermal broadening of the states as well as to the improvement of $T_{MG}$ at low temperature.

**Minimum conductivity and current on/off ratio**

The minimum conductivity of a graphene transistor which determines the "off-state" current has also been a widely studied topic[34-38]. It was predicted that for a perfect graphene sheet, its intrinsic minimum conductivity is determined by transport via contact-induced evanescent modes and approaches the value of $4e^2/\pi h$ at a large $W/L$ ratio[35]. In realistic devices, extra complexity is introduced by bulk disorder and the effect of contacts. In particular, the understanding of the role of contacts on minimum conductivity is still unclear[34]. Previous experimental studies suggested that for devices whose dimensions are smaller than the typical lengthscale of electron-hole puddles, size scaling of the minimum conductivity follows the evanescent mode model[36]. On the other hand, devices with larger dimensions can exhibit more complicated, finite size scaling effects[38]. In our experiment, all the devices have the same channel width of 1 μm with lithographically defined edges. Since our width is much larger than typical electron-hole puddles length scale[39], the puddle-induced minimum carrier density $n_{min}$ is not negligible. We plot $\sigma_{min}$ as a function of the aspect ratio $W/L$ in the inset of Fig. 3b. In the limit of a W/L ratio of about 20, it does approach $4e^2/\pi h$. However, the experimental trend clearly deviates from that predicted by the evanescent mode model shown as the black dashed curve in Fig. 3b. Instead, $\sigma_{min}$ can be well accounted for if we assume a propagating mode dominated transport model, where the number of modes $M = W \sqrt{\pi n_{min}} / \pi$ is given by the puddle-induced minimum carrier density. An effective transmission coefficient $T$ through the graphene channel and two graphene-metal junctions at the charge neutrality point is given by: $\dfrac{1-T}{T} = 2 \dfrac{1-T_{MG}}{T_{MG}} + \dfrac{L_{ch}}{\lambda_{mfp}}$. This leads to a quasi-ballistic formula for the minimum conductivity given by [28]:

$$\sigma_{min} = \frac{4e^2}{\pi h} L_{ch} \sqrt{\pi n_{min}} \left/ \left( \frac{L_{ch}}{\lambda_{mfp}} + \frac{2 - T_{MG}}{T_{MG}} \right) \right.$$, which is used to analyze the data in Fig. 3b.

The effectiveness of the modulation of a graphene transistor, especially for devices with very small channel length, is of vital importance to any related technological applications in analog electronics. Here we study the modulation ability in scaled graphene devices through their on/off current ratio for both n-type graphene transistors and p-type transistors as shown in Fig. 3c. Here the on/off ratio is defined as the ratio of maximum resistance at the charge neutrality point and the minimum of resistance at the p-side (-25 V away from $V_{CNP}$) or at the n-side (+25 V away from $V_{CNP}$). Notice that the on/off ratio of p-type transistors is larger than that of n-type transistors. This is mainly due to the electron hole asymmetry as discussed above. The on/off ratio decreases for smaller channel length devices, due to the transition of transport from diffusive to ballistic and the increasingly important role of the contact resistance. In the diffusive transport limit, the current is linearly dependent on the carrier density $n_{ch}$. However, in the ballistic regime, the current has a weaker carrier density dependence of $\sqrt{n_{ch}}$ . When the channel length is reduced below 10 nm, i.e. it becomes of the order of the barrier width of the heterojunction, the on/off ratio should approach unity due to the loss of the control of the Fermi-level in the graphene channel by the gate voltage. In a well-developed technology like that of silicon electronics, the continuous process of channel length down-scaling is accompanied by efforts to further reduce the contact resistance. In the case of graphene FETs, the lowest contact resistance achieved so far is around 100 ohm·μm, and there is no clear route to significantly decreasing this resistance. Unlike in conventional bulk semiconductors, metallic contacts not only introduce an extra resistor through $T_{MG}$, but also introduce a heterojunction, the chiral tunneling efficiency of which imposes extra resistance to transport. In order to achieve the maximum on/off ratio, one needs to choose a metal that would maximize this tunneling efficiency. The metal should have the right work function depending on the transistor polarity, namely, higher work function for p-type transistor and lower work function for n-type transistor. The current modulation results from the capability to modulate the number of conducting modes in the channel region (region III) and the electron-hole puddle density originating from disorder and scatterers, which could be improved with better quality graphene and insulator materials.

In summary, we have performed a systematic study of the behavior of graphene transistors upon changing the transport mechanism from diffusive to ballistic by varying the transistor channel length. Upon length scaling, clear signatures of electron quantum interference are seen. These are due to the metal contact-induced graphene heterojunction formation and indicate the appearance of the coherent transport regime. The nature of the electric current in such graphene heterostructures is discussed and attributed to the combined contributions from the metallic contact induced doping as well as the electrostatic control by the gate. The transport mechanism near the charge neutrality point has also been probed and propagating modes across electron-hole puddles formed are found to dominate the minimum conductivity.

**Methods**

We model the charge carrier density in the channel region III using:

$$n(x) = n_M \frac{1 - V_{ch}/V_M}{(x + x_b)/l_M} + n_{ch}(x) + n_M \frac{1 - V_{ch}/V_M}{(L_{ch} - x + x_b)/l_M} , \text{ for } 0 \leq x \leq L_{ch}, \qquad (1)$$

$en_{ch}(x) = -C_{BG}\left(V_D(x) + V_g\right)$, where the oxide thickness is $d = 90$ nm and the dielectric constant $\varepsilon_{oxide}$=3.9 determine the gate capacitance $C_{BG}$. At zero temperature and in the absence of disorder, the relationship between Fermi energy $eV_D(x)$ and the carrier density is given by $eV_D(x) = \text{sign}(n(x))\sqrt{n(x)/C_q}$, where the quantum capacitance is given by $C_q = \left(\pi \hbar^2 v_F^2\right)^{-1}$. The graphene metal doping $n_M(V_M)$ is calculated as in Ref. [32] (see supplemental information). The Fermi energy in the channel far away from the contacts $eV_{ch}$ is calculated using a carrier density $-C_{BG}V_g$. The values of $x_b$ are chosen to ensure a continuous electrostatic potential $V_D(x)$. Eq. (1) is motivated by the asymptotic form of the carrier density decay away from the metal contact in the large $d$ limit, see Ref. [18]. Note, that an effective $\kappa = (\varepsilon_{oxide} + \varepsilon_{air})/2$ defines the screening length that controls the p-n junction width[18,19] $l_M = 4\kappa\varepsilon_0\hbar v_F / \left(\sqrt{\pi n_M} e^2\right)$, which in turn determines the asymmetry and together with the channel length the effective cavity length, and, hence, the period of oscillations (see supplementary information). For small barrier width of the order of a few nanometers, we find that the detailed shape of the electrostatic potential has little impact on the results of the simulation. The p-n-p resistance (in the absence of broadening, zero temperature, and at zero bias) is given by $\dfrac{1}{R_{tot}} = \dfrac{4e^2}{hW} \displaystyle\sum_{n=0}^{M} T\left( k_y = \dfrac{\pi n}{W}, E_{x=0}, E_{x=L_{ch}} \right)$, where $E(x) = -eV_D(x)$. Extension to finite temperatures is straightforward[28].

**Acknowledgements**

The authors are grateful to B. Ek and J. Bucchignano for technical assistance and W. Zhu for discussions. The authors would also like to thank DARPA for partial financial support through the CERA program (contract FA8650-08-C-7838). The views, opinions, and/or findings contained in this article/presentation are those of the author/presenter and should not be interpreted as representing the official views or policies, either expressed or implied, of the Defense Advanced Research Projects Agency or the Department of Defense. Approved for Public Release, Distribution Unlimited.

**Figure Captions**

**Figure 1: Fabrication and ambipolar transfer characteristics for graphene transistors.**
**(a),** SEM image of an array of bottomed-gated graphene device with different channel length. Scale bar, 2 μm. Inset: SEM image of a 50 nm channel length device. Scale bar, 100 nm **(b)**, Schematic view of a bottom-gated graphene transistor and the transport processes in five regions. **(c)**, Total resistance versus back gate voltage sweep relative to charge neutrality point for three different channel lengths: 500 nm (black line), 170 nm (red line) and 50 nm (blue line) at room temperature and at 4.3 K **(d).**

**Figure 2: Experimental and theoretical analysis for ballistic short channel graphene devices. (a)** Resistance versus relative back gate voltage for a 50 nm device at 4.3K. Experimental data (blue curve) and ballistic modeling using: (1) a channel length $L_{ch}$=52 nm, Fermi level in graphene under the metal pinned at $V_0$=100 meV, and transparent metal to graphene junctions (red dashed line) and (2) using a non-ideal graphene metal barrier with $T_{MG}$ = 0.36, the Fermi level under the metal described by the electrostatic model (see Supplemental information) with a metal to graphene distance $d_1$ = 1.3 Å, and electron-hole puddles (see Supplementary information) with $n_{pd}$ = 1.1 × 10$^{12}$ cm$^{-2}$ , and $V_{pdM}$ =150 meV (red solid line). In both models (1) and (2) the resonance peak positions, marked by the vertical dashed lines, are determined by the cavity length, while the amplitude of the oscillations is determined by the reflection coefficient at the junction between graphene underneath the metal and graphene in the channel. In model (2), inhomogeneities reduce the oscillation amplitude and capacitive coupling to the metal gives rise to the gate dependence at negative $V_{BG}$. (**b**) Energy position of oscillation peaks from experiment (black), ballistic model 1 (red), and fit to the interferometer equation (blue). (**c**) Same as (**a**) at 300K: experiment (blue), model 1 (red-dashed), model 2 (solid red). (**d**) Resistance versus relative back gate voltage for a 70 nm device at 4.3 K. Experimental data (solid lines) and simulations (dashed lines) using model 2 (same parameter set as in (**a**)) at magnetic fields $B = 0$ T (black), 0.5 T (red), 1T (green), 1.5 T (blue) and 2 T (cyan). The arrows indicate the half-period shift under the magnetic field.

**Figure 3: Scaling behavior of resistance asymmetry and minimum conductivity for two terminal graphene devices.** (**a**) Resistance asymmetry versus channel length at room temperature (red circles) and at 4.3 K (blue squares). The lines are to guide the eye. (**b**) Experimental data of minimum conductivity versus channel length at room temperature (red circles), at 4.3 K (blue squares), and simulation of evanescent mode prediction (black dashed). For $T_{MG} \approx 0.4$, the fit to the propagating mode model (solid curves) uses a mean free path of ~ 75 nm and a minimum carrier density of about ~$3 \times 10^{11}$ cm$^{-2}$ Inset: The same data set versus aspect ratio W/L. (**c**) Current on/off ratio versus channel length at 4.3 K for n-FET (red) and p-FET (black). The lines are to guide the eye.

Figure 1

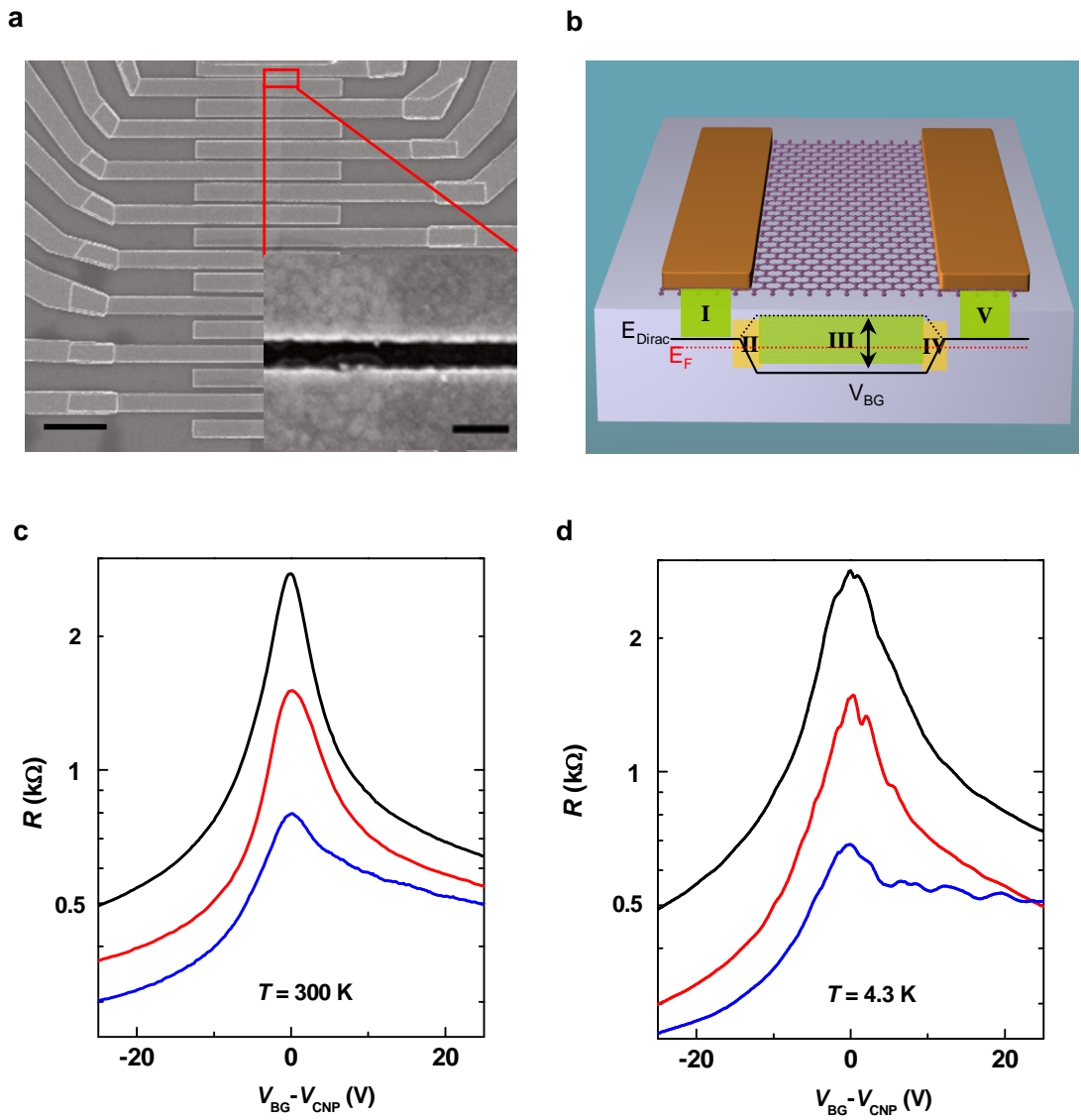

Figure 2

**a**

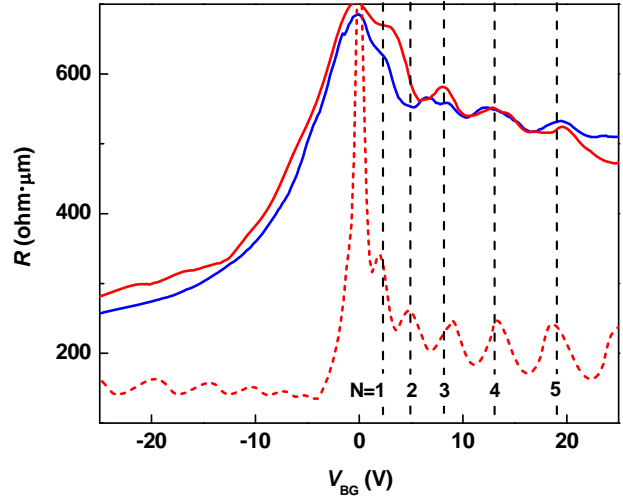

**b**

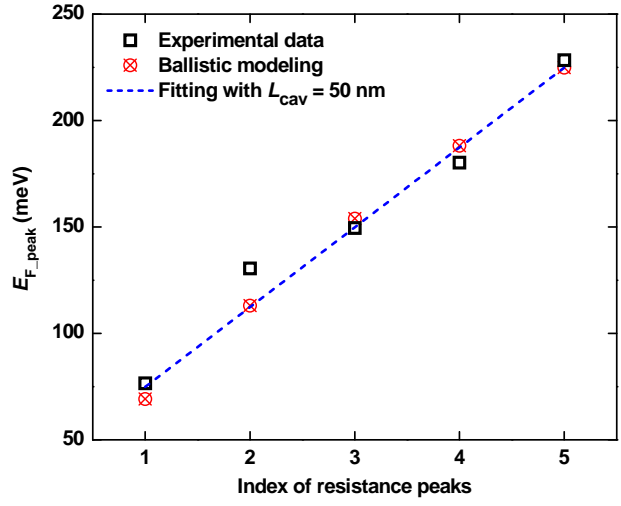

**c**

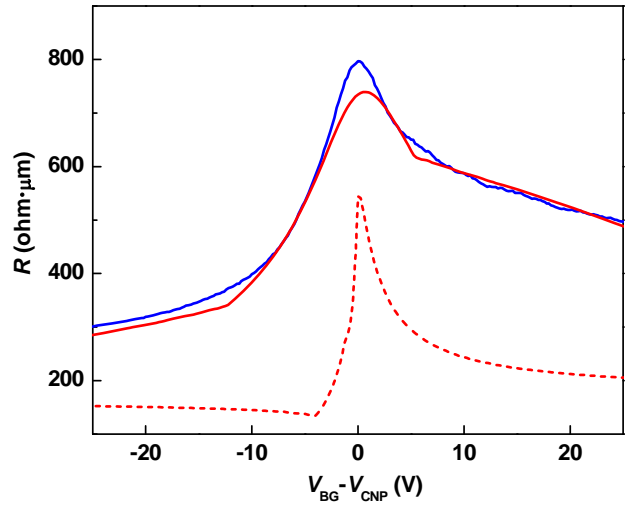

**d**

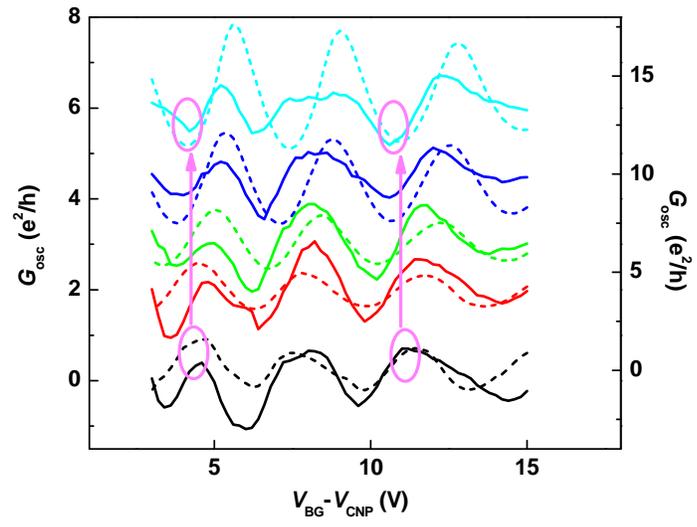

Figure 3

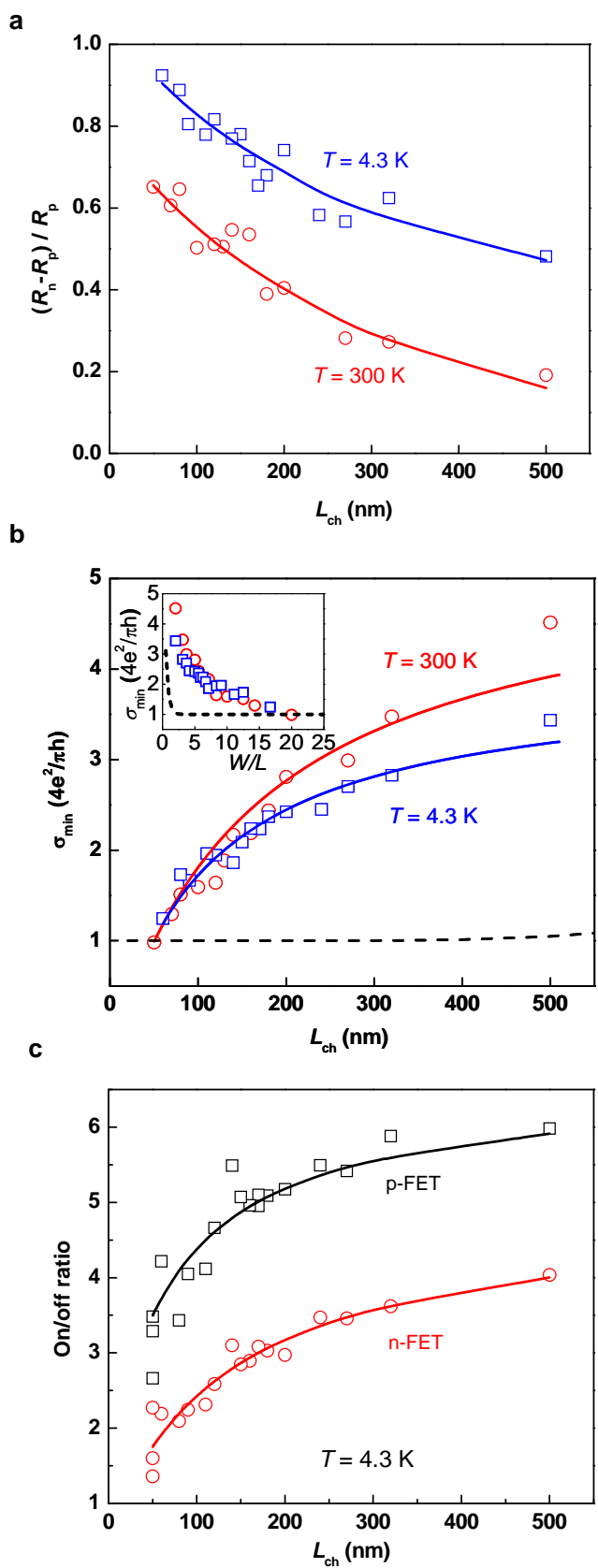